\documentclass[iop,tighten]{emulateapj}

\usepackage{subfigure}
\def\simgt{\lower 2pt \hbox{$\, \buildrel {\scriptstyle >}\over{\scriptstyle \sim}\,$}}
\def\simlt{\lower 2pt \hbox{$\, \buildrel {\scriptstyle <}\over{\scriptstyle \sim}\,$}}

\begin{document}

\title{On the detectability of star-planet interaction}

\author{Brendan~P.~Miller,$^{1}$ Elena~Gallo,$^{1}$
Jason~T.~Wright,$^{2,3}$ and Andrea~K.~Dupree$^{4}$}

\footnotetext[1]{Department of Astronomy, University of Michigan, Ann
  Arbor, MI 48109, USA}

\footnotetext[2]{Department of Astronomy and Astrophysics, The
  Pennsylvania State University, University Park, PA 16802, USA}

\footnotetext[3]{Center for Exoplanets and Habitable Worlds, The
  Pennsylvania State University, University Park, PA 16802, USA}

\footnotetext[4]{Harvard-Smithsonian Center for Astrophysics,
  Cambridge, MA 02138, USA}

\begin{abstract}

Magnetic (or tidal) interactions between ``hot Jupiters'' and their
host stars can potentially enhance chromospheric and coronal
activity. An ideal testbed for investigating this effect is provided
by the extreme WASP-18 system, which features a massive ($\sim$10
times Jupiter) close-in ($\simlt$1~day period) transiting planet
orbiting a young F6 star. Optical and \hbox{X-ray} observations of
WASP-18 were conducted in November 2011. The high-resolution echelle
spectrograph MIKE was used on the 6.5m Magellan Clay Telescope to
obtain 13 spectra spanning planetary orbital phases of 0.7--1.4, while
the X-ray Telescope on {\it Swift\/} provided contemporaneous
monitoring with a stacked exposure of $\sim$50~ks. The cores of the
\ion{Ca}{2} H and K lines do not show significant variability over
multiple orbits spanning $\sim$8~d, in contrast to the expectation of
phase-dependent chromospheric activity enhancements for efficient
star-planet interaction. The star is also \hbox{X-ray} faint, with
$\log{L_{\rm X}}<27.6$~erg~s$^{-1}$ (0.3--2~keV), indicating that
coronal activity is likewise low. The lack of detectable star-planet
interaction in this extreme system requires that any such effect must
here be transient, if indeed present. We demonstrate that searches for
\ion{Ca}{2} H and K variability can potentially mistake a stellar
hotspot, if observed over a short segment of the rotation period, for
planet-induced activity. Taken together, these results suggest that
the utility of star-planet interaction as a robust method of
estimating exoplanet magnetic field strengths may be limited.

\end{abstract}


\section{Introduction}

The possibility of magnetic star-planet interactions was initially
explored by Cuntz et al.~(2000; see also Rubenstein \& Schaefer~2000),
and this and subsequent work has indicated that reconnection events
could be an important effect in ``hot Jupiter'' systems, acting to
produce enhancements\footnote{Tidal star-planet interaction is also
possible, but likely relevant only in very close systems, if any, due
to the inverse-cube scaling with semi-major axis. Any tidal effect
should be maximal at planetary transit.} in chromospheric and coronal
activity co-rotating with the planet rather than with the stellar
rotation. It has been conjectured that the energy available from
reconnection events should scale with the product of the stellar and
planetary magnetic field strengths, as well as the relative velocity
of the magnetic field lines, and inversely with approximately the
square of the orbital semi-major axis (Cuntz et al.~2000; Kashyap et
al.~2008). Consequently, a calibrated relationship between the
amplitude of star-planet interaction and stellar and orbital
parameters could permit estimation of exoplanet magnetic field
strengths. Throughout this work we quantify interaction strength in
otherwise similar systems as $M_{\rm P}/a^{2}$, where $a$ is the
semi-major axis in AU and the planetary magnetic field strength is
taken to scale with the planetary mass $M_{\rm P}$ (e.g., Arge et
al.~1995; Stevens 2005); this is broadly consistent with the various
proposed trends investigated (e.g., Kashyap et al.~2008; Shkolnik et
al.~2008; Scharf 2010; Poppenhaeger et al.~2010). Numerical studies
suggest that star-planet interactions can potentially generate
sufficient energy to be observable as, e.g., phase-variable core
\ion{Ca}{2} H and K or \hbox{X-ray} emission (probing chromospheric
and coronal activity, respectively) in monitoring of individual hot
Jupiter systems, or as a greater average level of activity in systems
with more massive or closer-in planets. For example, Lanza et
al.~(2008, 2011) modeled chromospheric hot spots in several systems
(offset from the subplanetary point by varying degrees) as arising
from star-planet magnetic reconnection events, and Cohen et al.~(2009,
2011) carried out three-dimensional magnetohydrodynamic simulations
demonstrating that close-in giant planets can produce an increase in
overall stellar activity and generate (non-persistent) coronal hot
spots that rotate synchronously with the planet (albeit potentially
shifted in phase); see also Pillitteri et al.~(2010).

Observational evidence of magnetic star-planet interaction has now
been presented for several individual systems. For example, Shkolnik
et al.~(2005) examined 10 stars (K1 to F7) known at the time to
possess massive, close-in planets (median minimum planetary mass
$M_{\rm P}\sin{i}=0.6 M_{\rm Jup}$, median orbital period $P=3.4$~d),
in low-eccentricity orbits, and claimed evidence of star-planet
interaction in HD~179949 and $\upsilon$~And (both F8) based on slight
\ion{Ca}{2} H and K emission variability\footnote{The cores of the
\ion{Ca}{2} H and K lines are good indicators of chromospheric
activity; residuals from H$\alpha$ or the calcium infrared triplet are
also useful for this purpose.} synchronized with the planetary period
(see also discussion in Gu et al.~2005), although this synchronization
is apparently transient (Shkolnik et al.~2008). {\it Chandra\/}
observations of HD~179949 showed variable \hbox{X-ray} emission with
an apparent maximum near the phase associated with \ion{Ca}{2} H and K
enhancement (Saar et al.~2008). Fares et al.~(2012) carried out
spectropolarimetric observations of HD~179949, finding that the
stellar magnetosphere is highly tilted (producing two maxima per
rotation period) and that chromospheric activity is primarily linked
to stellar rotation, although low-level fluctuations near the beat
period could be planet-induced (see also Gurdemir et al.~2012). Later
{\it Chandra\/} observations of $\upsilon$~And (as well as concurrent
optical spectroscopy) did not find any indications of star-planet
interaction (Poppenhaeger et al.~2011a). Interaction was claimed for
$\tau$~Boo (F7) based primarily on photometric spot modeling, although
the tidal-locking of the star to the planet complicates interpretation
(Walker et al.~2008; see also Shkolnik et al.~2008). HD~189733 (K0)
twice showed an \hbox{X-ray} flare after secondary eclipse (Pillitteri
et al.~2010, 2011) but chromospheric variability appears to be tied to
stellar rotation (Shkolnik et al.~2008; Fares et al.~2010). Optical
spectroscopy of several systems (including $\upsilon$~And and
$\tau$~Boo) carried out by Lenz et al.~(2011) did not find evidence of
planet-induced chromospheric activity, but did identify potential
interaction between the M dwarf HD~41004B (which has binary,
planet-bearing K dwarf companion HD~41004A) and its brown-dwarf
companion HD~41004Bb. As a counterexample, 51~Peg (G5) is one of
several systems in which a sensitive search did not uncover evidence
of star-planet interaction; its low $L_{\rm X}$ and low
$\log{R^{'}_{\rm HK}}$ suggest it may in fact be in an extended state
of especially low activity, analogous to the Sun's Maunder minimum
(Poppenhaeger et al.~2009).

Several other studies have used large samples of planet-bearing stars
to test for significantly increased activity in hot Jupiter systems,
statistically averaging over orbital phase. \hbox{X-ray}
investigations of potential planet-induced enhancements in coronal
activity have to date provided mixed results. This may be due to the
difficulty of controlling for selection effects (more active stars,
which tend to be brighter in \hbox{X-rays}, are disproportionately
identified as planet-bearing with high mass, close-in planets, since
those stellar properties limit detectability of low mass, distant
planets) or may be related to sampling a heterogeneous mix of
late-type (typically F, G, K, and M) stellar classes. Kashyap et
al.~(2008) considered a large sample primarily drawn from {\it
ROSAT\/} pointed and all-sky survey (RASS) observations (containing
$\sim$30\%/70\% \hbox{X-ray} detections/upper limits) and found that,
after attempting to control for selection effects, stars with close-in
planets ($a<0.15$~AU) were 1.3--4 times more \hbox{X-ray} luminous
than stars with distant ($a>1.5$~AU) planets. Similarly,
Scharf~(2010), using RASS data again with a high percentage of upper
limits, did not find enhanced \hbox{X-ray} emission for close-in
planets (note that their distant planets encompassed $a>0.15$~AU) and
estimated their $L_{\rm X}-M_{\rm P}$ correlation to be robust against
observational bias. On the other hand, Poppenhaeger et al.~(2010)
looked at a volume-limited ($d<30$~pc) sample of stars observed with
{\it XMM-Newton\/} or {\it ROSAT\/} and found no significant
correlations of $L_{\rm X}/L_{\rm bol}$ with $M_{\rm P}$ or $a$; they
explain the possible correlation of $L_{\rm X}$ with $M_{\rm P}/a$ as
arising entirely from selection effects. Importantly, Poppenhaeger et
al.~(2011b) demonstrated that distance in shallow RASS datasets
covaries with both $L_{\rm X}$ and $M_{\rm P}$ (indicating that deeper
observations are essential to sidestep selection effects) and that
apparent correlations with $L_{\rm X}$ vanish when $L_{\rm X}/L_{\rm
bol}$ is used instead (thereby controlling, to first order, for
spectral class). Large-sample studies of potential planet-induced
enhancements in chromospheric activity have also given mixed results;
Canto Martins et al.~(2011) find no obvious correlations between
planetary mass or semi-major axis and the \ion{Ca}{2} H and K activity
indicator $\log{R^{'}_{\rm HK}}$, whereas Krej{\v c}ov{\'a} \& Budaj
(2012) do find these variables to correlate but only within the subset
of cooler host stars with $T_{\rm eff}<5500$~K.

In this paper, we present a sensitive search for star-planet
interaction in the extreme WASP-18 system. If star-planet interaction
is to become a useful probe of exoplanet magnetic field strengths in
hot Jupiter systems, additional convincing instances, beyond the
handful suggested to date, must be identified. Observations of a
strongly-interacting system could constitute a contextual template to
guide interpretation of results in more weakly-interacting systems
(including those already studied). Additionally, a robust measurement
of star-planet interaction in an extreme high-mass, short-period
system would supply productive leverage for uncovering and later
quantifying scaling relations. Conversely, a failure to detect
planet-induced stellar activity in an extreme system would severely
constrain the practical relevance of star-planet interaction and could
impact theoretical understanding and numerical modeling of this
effect.

\subsection{WASP-18 in context}

We identified WASP-18 as the potentially most strongly interacting
system among currently discovered exoplanets, a distinction it retains
as of April 2012. We obtained a list of all confirmed planets with
$M_{\rm P}>0.1M_{\rm J}$ (i.e., planetary mass greater than 10\% that
of Jupiter), orbital period $P<100$~d, and \hbox{$V<15$} from the
Exoplanet Orbit Database\footnote{Available at {\tt
http://www.exoplanets.org}} (Wright et al.~2011). As may be seen in
Figure~1, WASP-18b (Hellier et al.~2009; Southworth et al.~2009), with
$M_{\rm P}=10.4 M_{\rm J}$, $P=0.941$~d, and $a=0.020$~AU, has the
largest value of the interaction strength proxy $M_{\rm P}/a^{2}$
within this sample (and would also rank first for other plausible
parameterizations of interaction strength). More specifically, after
excluding brown dwarfs (with $M_{\rm P}>13 M_{\rm Jup}$), of the 551
entries in the Exoplanet Orbit Database, the $M_{\rm
P}/a^{2}\simeq25000$~$M_{\rm Jup}$~AU$^{-2}$ for WASP-18 is nearly 2.5
times greater than the next largest value (CoRoT-14b), and only 5
systems have $M_{\rm P}/a^{2}>5000$~$M_{\rm Jup}$~AU$^{-2}$. Because
WASP-18 is a transiting system, the planetary properties are securely
established by lightcurve and radial-velocity measurements. The values
of $M_{\rm P}/a^{2}$ are also labeled for several systems described in
$\S$1; for example, HD~179949 has $M_{\rm P}/a^{2}\simeq500$~$M_{\rm
Jup}$~AU$^{-2}$.

\begin{figure*}
\includegraphics[scale=0.78]{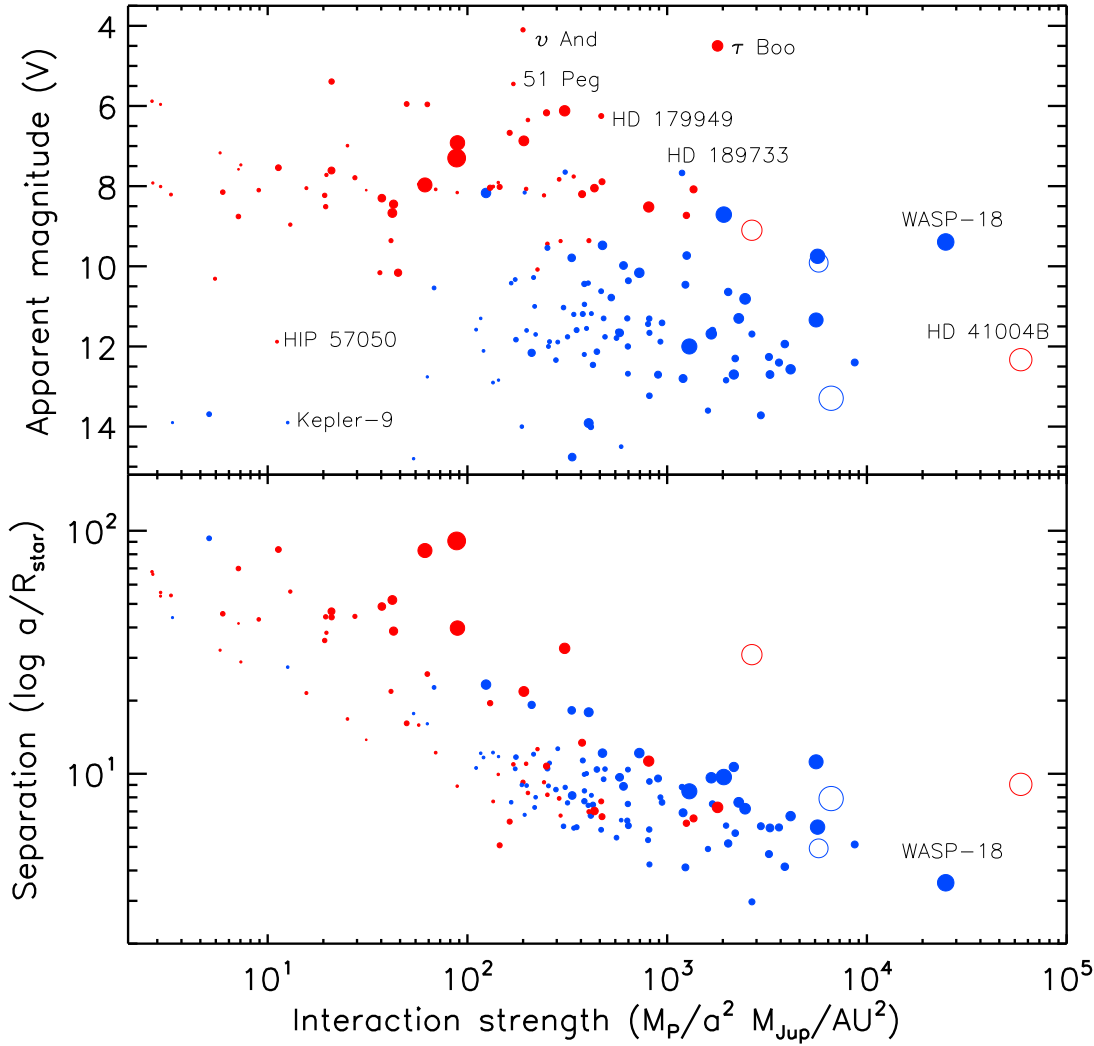} \figcaption{\small
Distribution of star-planet interaction strength, assumed here to
scale as $M_{\rm P}/a^{2}$, plotted versus apparent $V$ magnitude
(top) and star-planet separation in units of stellar radii
(bottom). Note that the horizontal axis is logarithmic. Transiting
planets are plotted in blue, and symbol size scales as $M_{\rm
P}^{0.5}$. Brown dwarfs ($>13M_{\rm J}$) are plotted as open
symbols. Radial-velocity masses are lower limits, but establishing
$\sin{i}$ would typically result in only modest rightward movement of
those points. Several relevant or interesting systems are labeled (the
$b$ planet designation is implicit).}
\end{figure*}

The stellar and planetary properties for the WASP-18 system are given
in Table~1, as are those for previously identified star-planet
interaction candidates HD~179949, $\upsilon$ And, and $\tau$ Boo (but
recall $\tau$~Boo is tidally locked, with stellar and orbital periods
of $\sim$3.2~d, which may suppress magnetic interaction). The value of
$M_{\rm P}/a^{2}$ for WASP-18b exceeds that for the other candidate
systems by 1--2 orders of magnitude. Such extreme planetary systems
are difficult to maintain, and WASP-18b itself is expected to have a
short lifetime against infall from tidal drag (Hellier et
al.~2009). The star WASP-18 has spectral type F6, effective
temperature $T_{\rm eff}=6400\pm100$~K, and stellar mass $M_{\rm
*}=1.22\pm0.03$~$M_{\odot}$ (values from the Exoplanet Orbit Database,
from which original references may be obtained); conveniently, these
are similar to the properties of HD~179949 (F8, 6170~K,
1.18~$M_{\odot}$), $\upsilon$ And (F8, 6210~K, 1.31~$M_{\odot}$), and
$\tau$ Boo (F7, 6390~K, 1.34~$M_{\odot}$). WASP-18 is slightly hotter
than HD~179949 and $\upsilon$~And, and consequently likely possesses a
slightly shallower outer convection zone. On the other hand, the
stellar rotation period is shorter, at $\sim$5.6~d, estimated from
$v\sin{i}=11$~km~s$^{-1}$; the orbit and stellar rotation are
aligned\footnote{All four of these systems are near the $T_{\rm
eff}\sim6250$~K border above which hot-Jupiter hosts tend to have high
stellar obliquities (Winn et al.~2010), but the spin-orbit alignment
in WASP-18 is more characteristic of cooler hot-Jupiter hosts.}
(Triaud et al.~2010). HD~179949 and $\upsilon$~And have rotational
periods of 7.6~d (Fares et al.~2012) and $\sim$12~d (Shkolnik et
al.~2008), respectively. Further, WASP-18 appears to be a younger
star, with an age of $\sim$500-700~Myr (Hellier et al.~2009; Brown et
al.~2011). The age-rotation-activity relation would then predict
relatively greater intrinsic activity in WASP-18, apart from any
planet-induced modulation or enhancement (but see $\S$3.2 for
caveats). Observationally, WASP-18 has other appealing features: the
star is among the brighter transit-detected systems, and the short
planetary period facilitates rapid accumulation of phase coverage
across multiple orbits. Due to its extreme properties, WASP-18 has
been studied at a range of frequencies; for example, Nymeyer et
al.~(2011) used {\it Spitzer\/} observations of secondary eclipse to
infer that the planet has $T\simeq$3100~K with near-zero values for
both albedo and day/night side energy redistribution. However,
high-resolution, high signal-to-noise \ion{Ca}{2} H and K spectroscopy
and sensitive \hbox{X-ray} observations have not been published prior
to the observations presented here.

\section{Observations}

\subsection{Data acquisition and reduction}

Optical spectroscopy was carried out with the 6.5m Clay Telescope on
Nov.~6--7 and 10--13 2011 using the Magellan Inamori Kyocera Echelle
(MIKE), a high-throughput double echelle spectrograph. A
$0.7{\times}5''$ slit was used throughout, providing a resolution of
R$\simeq$40000 near the \ion{Ca}{2} H and K lines. A communication
board failure prevented use of the blue-side CCD on Nov.~10 and bad
weather limited observing on Nov.~11. In total, 13 visits to WASP-18
were obtained, with combined exposure times of 20--40 minutes per
visit. The planetary orbital phase coverage (Figure~2) spans 0.7--1.4
and includes one visit at central planetary transit.

The spectra were reduced using the IDL MIKE Redux\footnote{{\tt
http://web.mit.edu/$\sim$burles/www/MIKE/}} package. Briefly, milky
flats (taken of $\delta$ Ori with the diffuser) were combined and
normalized to correct pixel response variations; trace flats (from
internal lamps) were combined for order and slit tracing; arc images
(including one associated with each science exposure) were processed
and used to derive two-dimensional wavelength maps, using a
fifth-degree polynomial fit along the orders which was found to give
generally structureless residuals; the slit profile was determined for
each order; science exposures were processed (bias-subtracted from
overscan and flat-corrected), the object was traced and the sky
subtracted, and the object flux was optimally extracted. Flux is
retained in relative units (i.e., no spectrophotometric standard star
calibration was conducted) as we are interested in normalized
spectra. Automated cosmic ray correction was not performed due to
concern about potential introduction of spurious variability; two
science frames with an obvious cosmic ray located near the
two-dimensional center of the \ion{Ca}{2} K line were discarded.

Contemporaneous monitoring was obtained with the {\it Swift\/}
satellite from Nov.~1--17 (Target ID 32149), notably with the
\hbox{X-ray} Telescope (XRT) through 58 snapshots of 500-1500 seconds
each.The XRT observations were reduced using {\it HEASARC\/} version
6.11.1. An exposure map was generated for each sequence, including a
vignetting correction and with time-dependent instrument maps used
whenever attitude variations exceeded 2.4$''$. Images were extracted
in both the full (0.3--10~keV) and soft (0.3--2~keV) bands. The
stacked XRT effective exposure from all sequences at the position of
WASP-18 reached $\sim$50~ks. The UV/Optical Telescope (UVOT) was also
used to obtain both photometry (with the U, UVW1, UVM2, and UVW2
filters) and UV grism spectroscopy, with coverage of the Mg~II h and k
lines which are also chromospheric activity indicators (Buccino \&
Mauas 2008).

\begin{figure}
\includegraphics[scale=0.45]{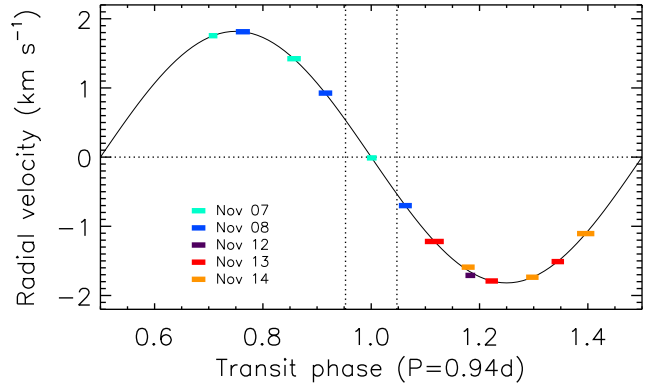} \figcaption{\small Phase coverage
of the Magellan observations, plotted on the stellar radial velocity
curve from the ephemeris of Hellier et al.~(2009; error here
$<0.001$). Dates are UT 2011; phase coverage of the 0.94~d planetary
orbit was accumulated over five nights. The Nov 12 and first Nov 14
points have slight vertical offsets for clarity. The vertical dotted
lines mark planetary transit. }
\end{figure}

\subsection{Analysis and results}

We checked for phase-dependent variability, as from one-sided enhanced
chromospheric activity, within the cores of the \ion{Ca}{2} H and K
lines, focusing primarily upon the latter. The \ion{Ca}{2} K line and
nearby stellar continuum were isolated through selecting wavelengths
spanning 3920--3950\AA, which included contributions from two
orders. For each visit, the flux from individual exposures was summed,
and the wavelength scale was adjusted through subtraction of a
center-of-mass velocity of 3.1961 km~s$^{-1}$ and a stellar reflex
velocity with an amplitude of 1.8183 km~s$^{-1}$, using the values and
ephemeris from Hellier et al.~(2009). Continuum normalization near the
\ion{Ca}{2} H and K regions is difficult because the absorption lines
are quite broad relative to the blaze function, there are a large
number of nearby metal lines, and the Wien tail is falling off
steeply. We used a similar normalization method to that conducted and
detailed by Shkolnik et al.~(2005, 2008), removing a slight linear
trend over $\simeq$7~\AA~centered near the line cores, such that unity
represents approximately one-third of the local stellar continuum. The
normalized \ion{Ca}{2} H and K spectra for all 13 visits are plotted
in Figure~3 (top).

\begin{figure*}
\includegraphics[scale=0.97]{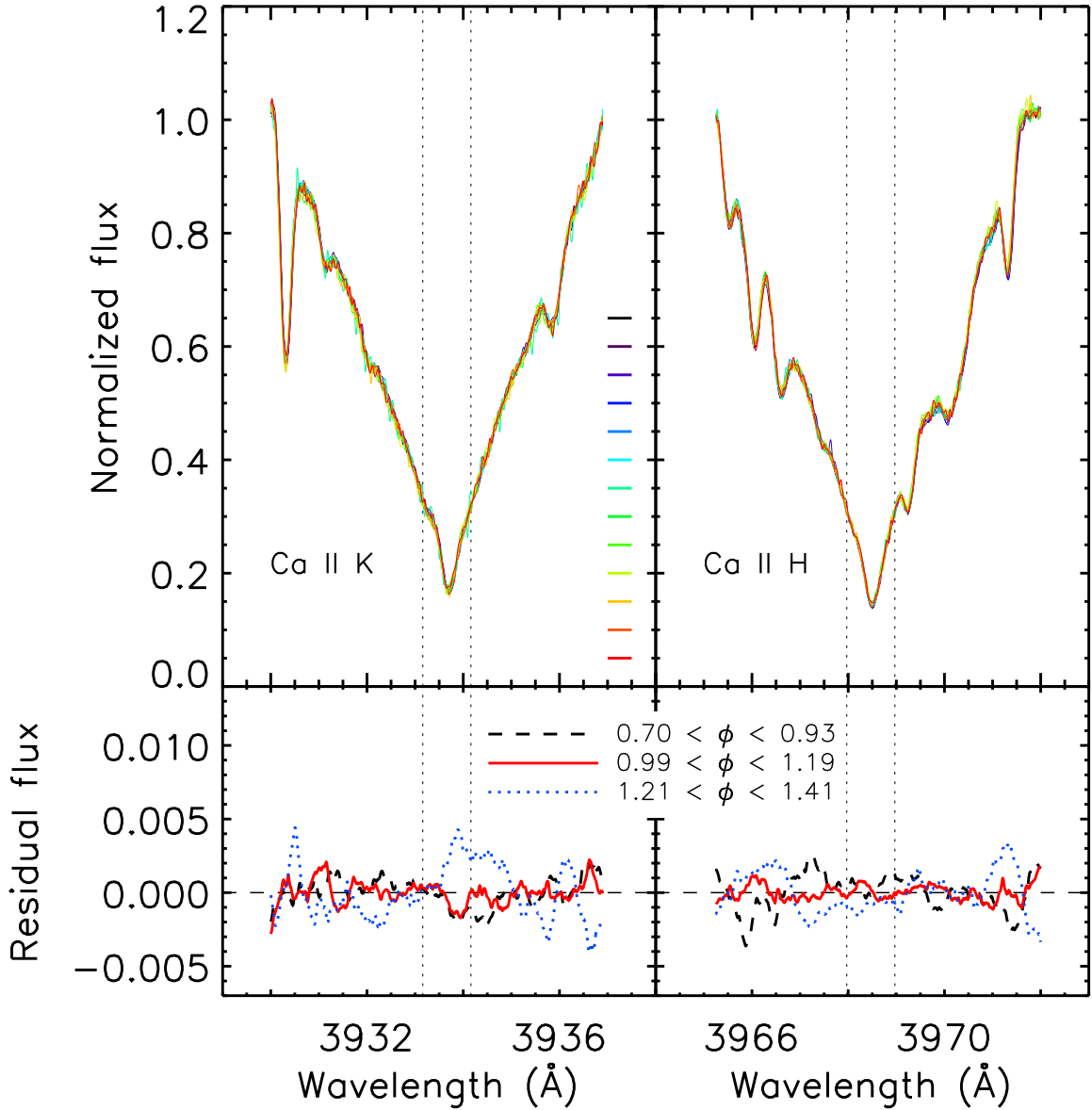} \figcaption{\small
Normalized spectra of the \ion{Ca}{2} H (right, top) and K (left, top)
regions overplotted for each of 13 visits. There is no obvious change
in the depth of the central core of the \ion{Ca}{2} K line. Residuals
from the median spectrum, binned by planetary orbital phase and boxcar
smoothed with a width of 0.4~\AA, are shown at bottom (after removal
of slight linear trends). The vertical dotted lines mark the
\ion{Ca}{2} H and K cores.}
\end{figure*}

Residuals from median normalized \ion{Ca}{2} H and K spectra were
calculated for each visit, and then stacked (weighted by exposure
times corrected for seeing) within orbital phase bins of 0.70--0.93,
0.99--1.19, and 1.21--1.41; these bins then contain spectra from 4, 5,
and 4 visits, respectively. The residuals were boxcar smoothed with a
width of 0.4\AA~and then any remaining slight linear trend was
subtracted to avoid influencing comparison of the line core. There is
no obvious difference in the residual spectra within the three
separate phase bins, to an amplitude of 0.003 of the normalized
continuum (Figure~3, bottom). Integrating the normalized flux over the
1~\AA~line cores within phase bins of 0.70--0.93, 0.99--1.19, and
1.21--1.41 gives nearly identical values: 0.252$\pm$0.001,
0.251$\pm$0.002, and 0.253$\pm$0.003 for K and 0.225$\pm$0.001,
0.227$\pm$0.001, and 0.228$\pm$0.003 for H, where the 1$\sigma$
uncertainties have been estimated as the standard deviation of the
component flux measurements within each phase bin. The 3$\sigma$ limit
on the relative flux variability is $<$3.2\% and $<$3.4\% for the K
and H line cores, respectively. The apparent marginal tendency for the
1.21--1.41 K bin to show relatively more red-side emission is within
the noise level\footnote{This may be verified through comparison to
the residuals near photospheric absorption features, which should not
show any variability phased with the planetary orbit.} but in any case
this slight profile variation is not present in H, is lopsided rather
than symmetric about the core, and lags rather than leads the
subplanetary point; these effects are inconsistent with planetary
magnetic interaction.

The XRT observations were initially planned for similar phase-resolved
analysis, but the star proved to be unexpectedly \hbox{X-ray} faint
(see also $\S$3.2) and so all sequences were combined for an effective
exposure of $\sim$50~ks. WASP-18 is not detected by XRT in the stacked
exposure, in either the soft (Figure~4) or the full
bands. Specifically, within an aperture of 20$''$ centered on the
SIMBAD optical position of WASP-18, there are 3.8 counts in the
0.3--2~keV band, where the expected background is 3.7 counts. The 95\%
upper limit on the source counts within 20$''$ is $<5.9$ net counts
(from Kraft et al.~1991), corresponding to $<8.1$ total net counts
after accounting for the XRT point spread function (Moretti et
al.~2005). For a coronal model with $kT$=1~keV and solar abundances
the unabsorbed 0.3--2 keV \hbox{X-ray} flux is
$<3.3{\times}10^{-15}$~erg~s$^{-1}$, calculated using the Portable
Interactive Multi-Mission Simulator (PIMMS\footnote{{\tt
http://cxc.harvard.edu/toolkit/pimms.jsp}}) for a plasma/APEC model
presuming an intervening column of $N_{\rm H}=10^{18-19}$~cm$^{-2}$;
other plausible models give similar results to within $\simlt$30\% (or
0.1 dex). At the Hipparcos distance to WASP-18 of $100\pm10.6$~pc,
this corresponds to a limiting \hbox{X-ray} luminosity of $\log{L_{\rm
X}}<27.6$~erg~s$^{-1}$ (0.3--2~keV).

\section{Discussion}

We compare the observed persistent low activity in WASP-18 to
expectations for planet-induced variability, as well as to the
intrinsic properties of similar stars, and additionally explore the
general possibility of stellar hotspots acting to mimic star-planet
interaction.

\subsection{Expected variability in WASP-18}

Past work would seem to suggest that the degree of planet-induced
\ion{Ca}{2} H and K variability expected for WASP-18 could be
substantial. For example, Shkolnik et al.~(2005, 2008) report
``on/off'' \ion{Ca}{2} H and K variability phased with the planetary
orbital period in HD~179949 and $\upsilon$~And (see also Poppenhaeger
et al.~2011a), which have stellar types of F8 and do not show obvious
core emission within the deep photospheric absorption (similar to
WASP-18). The amplitude of residual flux associated by Shkolnik et
al.~(2008, Figure 3; 2005, Figure 6) with star-planet activity in
HD~179949 ($\upsilon$~And) is $\simeq$0.017
($\simeq$0.008).\footnote{The ratio of residuals for HD~179949 and
$\upsilon$~And is $\sim$2, while the ratio of $M_{\rm P}/a^{2}$ is
$\sim$2.5. However, given the dissimilarity of other parameters, such
as intrinsic stellar activity, and the apparent variability of any
star-planet interaction in these systems, we do not consider this
necessarily informative.}  The value of $M_{\rm P}/a^{2}$ for WASP-18
could suggest an effect 50 (130) times greater than for HD~179949
($\upsilon$~And).\footnote{While less physically motivated,
alternative scalings with either separately $1/a^{2}$ or $M_{\rm P}$
would still predict a substantially greater effect for WASP-18, by a
factor of 4.8 (8.7) or 11 (15) compared to HD~179949
($\upsilon$~And).} In contrast to such predictions, the level of
phase-binned variability observed for WASP-18 is $\simlt$0.003
(Figure~3; $\S$2.2), below the levels detected in HD~179949 and
$\upsilon$~And.

The degree of \hbox{X-ray} variability expected in WASP-18 is more
difficult to estimate, as only a handful of \hbox{X-ray} brightening
events have been attributed to potential star-planet interaction. For
example, HD~179949 displayed an increase in \hbox{X-ray} emission by
$\sim$30\% near the phase associated with the \ion{Ca}{2} H and K
enhancement (Saar et al.~2008); with a stellar \hbox{X-ray} luminosity
of $\log{L_{\rm X}}=28.6$~erg~s$^{-1}$ (Kashyap et al.~2008), this is
an increase of $\log{L_{\rm X}}=28.1$~erg~s$^{-1}$. HD~189733 showed
two \hbox{X-ray} flares near phase 0.53 (i.e., shortly after
occultation), peaking at twice the baseline count rate (Pillitteri et
al.~2011); with a stellar \hbox{X-ray} luminosity of $\log{L_{\rm
X}}=28.4$~erg~s$^{-1}$ (Kashyap et al.~2008), this is an increase of
$\log{L_{\rm X}}=28.4$~erg~s$^{-1}$. Cohen et al.~(2011) used their
MHD modeling of the HD~189733 system to estimate that the energy
available from magnetic reconnection to accelerate particles into the
stellar corona is $\sim$10$^{28}$~erg~s$^{-1}$ (after applying
conservative efficiency assumptions). From these examples, WASP-18
might be expected, taking into account its exceptional value of
$M_{\rm P}/a^{2}$, to possess interaction energy available for coronal
activity enhancement sufficient to generate absolute increases in its
\hbox{X-ray} luminosity by several times
$10^{28}$~erg~s$^{-1}$. However, the observed XRT upper limit of
$\log{L_{\rm X}}<27.6$~erg~s$^{-1}$ indicates that the \hbox{X-ray}
luminosity did not achieve this level for any more than $\simlt$10\%
of the {\it Swift\/} coverage. Without even a stacked \hbox{X-ray}
detection it is not possible to evaluate the relative amplitude of
phase-dependent \hbox{X-ray} variability, if any, in WASP-18.

\begin{figure}
\includegraphics[scale=0.45]{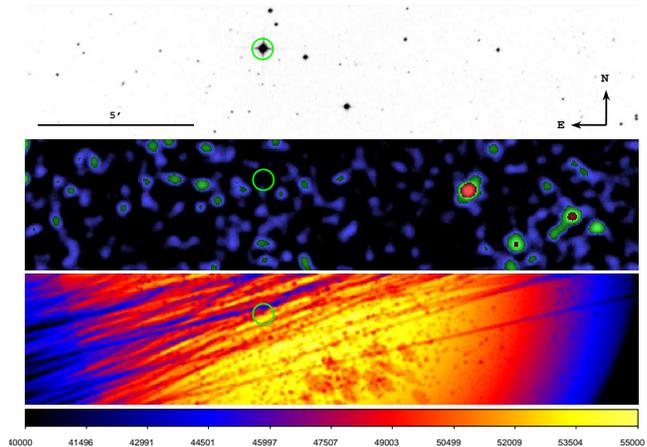} \figcaption{\small {\it
Top:\/} Digital Sky Survey image of WASP 18. {\it Middle:\/} {\it
Swift\/} XRT 0.3--2~keV image of WASP 18 region, Gaussian smoothed
with a kernel of 7 pixels and shown with logarithmic scaling. The
green circle has a radius of $20''$ and is centered on the SIMBAD
position of WASP 18. {\it Bottom:\/} XRT exposure map; the color bar
gives the effective exposure in seconds.}
\end{figure}

\begin{deluxetable*}{p{40pt}rrrrrrrrrrrr}
\tablecaption{Candidates for star-planet interaction}
\tabletypesize{\footnotesize}
\tablewidth{17.7cm}

\tablehead{ & \multicolumn{4}{c}{Star} & \multicolumn{3}{c}{Planet} &
\multicolumn{5}{c}{Activity} \\ \colhead{Name} & \colhead{Type} &
\colhead{$V$} & \colhead{$B$$-$$V$} & \colhead{$d$} & \colhead{$M_{\rm
P}$} & \colhead{$P$} & \colhead{$a$} & \colhead{Age} &
\colhead{$M_{P}/a^{2}$} & \colhead{$R_{\rm HK}^{'}$} &
\colhead{$\log{L_{\rm X}}$} & \colhead{$\log{(\frac{L_{\rm X}}{L_{\rm
bol}})}$} \\ & & & & \colhead{(pc)} & \colhead{($M_{\rm J}$)} &
\colhead{(d)} & \colhead{(AU)} & \colhead{(Gyr)} & \colhead{($M_{\rm
J}/$AU$^{2}$)} & & \colhead{(erg~s$^{-1}$)} & }

\startdata
WASP$-$18         &  F6  &   9.39 & 0.49 &  100   &  10.4  &    0.94 &   0.0201 &  0.6 &    25700 &   $<$$-$5.0\tablenotemark{a}  &  $<$27.6 &$<$$-$6.2  \\
$\tau$~Boo\tablenotemark{b}   &  F7  &   4.50 & 0.51 &  15.62 &   4.1  &    3.31 &   0.0480 &  2.5 &     1780 &   $-$4.73  &     28.8 &   $-$5.3  \\
HD~179949         &  F8  &   6.25 & 0.55 &  27.5  &  0.95  &    3.09 &   0.0439 &  2.1 &      490 &   $-$4.80  &     28.6 &   $-$5.3  \\
$\upsilon$~And    &  F8  &   4.10 & 0.54 &  13.49 &  0.69  &    4.62 &   0.0594 &  3.8 &      200 &   $-$5.07  &     27.6 &   $-$6.5  \\[-15pt]
\enddata

\tablecomments{Data is taken from the Exoplanet Orbit Database ({\tt
http://www.exoplanets.org}), except for planetary mass and system age
which are from the Extrasolar Planets Encyclopedia ({\tt
http://www.exoplanet.eu}), and $\log{L_{\rm X}}$ which is from this
work, Poppenhaeger et al.~(2012), Kashyap et al.~(2008), and
Poppenhaeger et al.~(2010) for WASP-18, $\tau$~Boo, HD~179949, and
$\upsilon$~And, respectively; $\log{(\frac{L_{\rm X}}{L_{\rm bol}})}$
is calculated. }  

\tablenotetext{a}{Estimated from our measured $\log{L_{\rm
X}}<27.5$~erg~s$^{-1}$ and the $R_{\rm HK}^{'}-\log{(L_{\rm X}/L_{\rm
bol})}$ correlation presented in Mamajek \& Hillenbrand (2008). See also
Footnote~15.}

\tablenotetext{b}{The potential strength of magnetic
star-planet interaction in the $\tau$~Boo system is reduced due to the
apparent tidal locking of the star to the planet; see $\S$1.}
\end{deluxetable*}

The scaling of star-planet interaction strength with $M_{\rm P}/a^{2}$
that we (and others; $\S$1) have adopted approximately holds for
otherwise similar systems. We briefly consider the impact upon the
energy available for star-planet interaction due to potentially
differing planetary or stellar magnetic field strengths in WASP-18
compared to HD~179949 and $\upsilon$ And. The planetary magnetic field
strength depends most relevantly upon $M_{\rm P}$ but also scales
inversely with the planetary rotational period. For tidally-locked
planets, as all these are believed to be (out to $\sim$0.15~AU;
Bodenheimer et al.~2001), the planetary rotation period is identical
to the orbital period, and both are substantially shorter for WASP-18
than for HD~179949 or $\upsilon$ And. This suggests (Arge et al.~1995;
S{\'a}nchez-Lavega 2004; Stevens 2005) an increase in baseline
planetary magnetic field strength by a factor of a few, which would
enhance any star-planet interaction. The relative velocity between
magnetic field lines, which is governed by the difference between the
planetary and stellar rotation rates, is $v_{\rm rel}=K(R_{\rm
*}/a)-v_{\rm rot}$ for tidally-locked planets, where $K$ is the
orbital velocity and $v_{\rm rot}$ is the equatorial stellar
rotational velocity (Cuntz et al.~2000). This would suggest a further
increase in interaction strength by a factor of a few. On the other
hand, the interaction strength approximately scales with the stellar
magnetic field, $B_{\rm *}$ (Cuntz et al.~2000; Kashyap et al.~2008;
see Lanza 2009 and Scharf 2010 for slightly different formulations),
which appears to be substantially weaker in WASP-18 than in at least
HD~179949 and perhaps also $\upsilon$ And. This is evident by the
\hbox{X-ray} non-detection of WASP-18, as $B_{\rm *}$ is observed to
depend approximately linearly upon $L_{\rm X}$ (Pevtsov et
al.~2003). That $\log{(L_{\rm X}/L_{\rm bol})}$ is at least one order
of magnitude lower in WASP-18 than HD~179949 possibly offsets any
gains in interaction strength due to faster planetary rotation,
although the difference in inferred stellar magnetic field strengths
is less between WASP-18 and $\upsilon$ And. However, since the
observed amplitude of \ion{Ca}{2} H and K variability in WASP-18 is
already $\simgt$5 times lower than that suggested to arise from
star-planet interaction in HD~179949, $B_{\rm *}$ would need to be
$\simgt$50 times lower in WASP-18 than in HD~179949 to explain our
non-detection for a similar interaction efficiency, which seems
implausible given their similar stellar spectral types.

The lack of evidence for star-planet interaction in WASP-18 is
unexpected given that we selected it as the system with the greatest
predicted interaction strength (based on planetary mass and semi-major
axis) and given the broad similarity in stellar properties between
WASP-18 and other systems (notably HD~179949 and $\upsilon$ And) for
which star-planet interaction has been claimed. Star-planet
interaction in WASP-18 would seem to be at best highly transient as it
was not demonstrably present during our observations. For reference,
\ion{Ca}{2} H and K variability has been described as phased with the
planetary orbit in $\simlt$50\% of the epochs at which HD~179949 and
$\upsilon$~And have been observed. Further, the timescale for a
magnetic reconnection event in the simulations of Cohen et al.~(2011)
is short, $\sim$0.25 orbits. On the other hand, our spectroscopic
coverage of WASP-18 extends over 8 complete planetary orbits, and the
\hbox{X-ray} observations cover many more (albeit at lower sensitivity
per orbit). To the extent that star-planet interactions are transient,
they should intuitively occur with greater frequency in extreme
systems such as WASP-18, which contrasts (if not definitively) with
our findings. It might be alternatively suggested that, if our
observations did not simply happen to take place within a period of
relative quiescence, the efficiency of star-planet interaction for
WASP-18 could be lower than has been identified in the past for less
extreme systems. However, we emphasize again that WASP-18 is not
notably distinctive in terms of stellar properties from HD~179949 and
$\upsilon$~And. (Below, we explore the possibility that the
particularly massive and close-in planet in WASP-18 actually acts to
suppress stellar activity.)

Regardless of its underlying cause, this lack of observed variability
in WASP-18 demonstrates that even extreme systems, arguably the best
candidates to display planet-induced activity enhancements, challenge
prevailing ideas concerning star-planet interaction. The appealing
prospect of calibrating star-planet interactions to estimate exoplanet
magnetic field strengths would currently appear to require additional
unambiguous evidence of such interactions occurring.

\subsection{Expected intrinsic activity in WASP-18}

We next consider the overall baseline level of chromospheric and
coronal activity in WASP-18, to provide context for the non-detection
of planet-induced variability. WASP-18 shows atypically low
\ion{Ca}{2} H and K core emission for its spectral class and inferred
age (contrast with, e.g., HD 111456; Freire Ferrero et al.~2004). When
stellar age is calculated from \ion{Ca}{2} activity, main-sequence F,
G, K, or M stars with inferred ages of 340--740~Myr (i.e., similar to
the age of WASP-18, which is not based on \ion{Ca}{2}
activity\footnote{The age of WASP-18 is estimated as
$630^{+950}_{-530}$~Myr by Hellier et al.~(2009) based on stellar
isochrones, and as $579^{+305}_{-250}$~Myr by Brown et al.~(2011)
based on tidal interaction modeling.}) are chromospherically
``active'' (following the definition of Henry et al.~1996), with
characteristic values of $-4.5<R_{\rm HK}^{'}<-4.4$ (from data in
Wright et al.~2004). While stars with $0.48<B-V<0.52$, as for WASP-18,
display a wide range of activity, they are significantly less likely
to have $R_{\rm HK}^{'}<-5.0$ (19\%, versus 37\% for F, G, K, and M
stars; from data in Wright et al.~2004). However, $R_{\rm HK}^{'}$ in
WASP-18 is low; if estimated indirectly based on the $R_{\rm
HK}^{'}-\log{(L_{\rm X}/L_{\rm bol})}$ correlation presented in
Mamajek \& Hillenbrand (2008), the measured \hbox{X-ray} limit for
WASP-18 suggests $R_{\rm HK}^{'}\simlt-5.0$.\footnote{A recent high
signal-to-noise Keck/HIRES spectrum of WASP-18, calibrated as
described in Wright et al.~(2004), gives $R_{\rm HK}^{'}=-5.15$
(Howard Isaacson 2012, private communication).}

The \hbox{X-ray} luminosity for WASP-18 is also lower than predicted.
Nine nearby stars of spectral class F5V--F7V (i.e., matched to
WASP-18) in a volume-limited {\it ROSAT\/} sample have \hbox{X-ray}
luminosities of $27.5<\log{L_{\rm X}}<29.0$~erg~s$^{-1}$, with mean
and median values of $\log{L_{\rm X}}\simeq28.3$~erg~s$^{-1}$ (from
data in Schmitt \& Liefke 2004). However, WASP-18 is younger than
typical main-sequence stars and should consequently be more
\hbox{X-ray} luminous. While we are not aware of a specific
comprehensive study of young F stars, in general the \hbox{X-ray}
luminosity of F5$+$ stars modestly exceeds that of G stars (Schmitt
1997) and Hyades G stars (hence of a similar age to WASP-18) have mean
$\log{L_{\rm X}}\simeq28.8$~erg~s$^{-1}$ (Stelzer \&
Neuh{\"a}user~2001; Preibisch \& Feigelson 2005). Consequently, the
XRT-derived limit of $\log{L_{\rm X}}\simlt27.6$~erg~s$^{-1}$ marks
WASP-18 as notably \hbox{X-ray}-faint relative to comparable stars,
with a $\log{(L_{\rm X}/L_{\rm bol})}\simlt-6.2$ similar to that of
the Sun. 

In summary, not only is there no apparent enhancement in chromospheric
or coronal activity in WASP-18 that might be linked to interaction
with the planet, but rather the above results, taken together, suggest
that WASP-18 is unusually quiet for a young F6 star. It is possible
that the stellar age is incorrect; the 1$\sigma$ error bars on the
measurement given by Hellier et al.~(2009) encompass $\simgt$1~Gyr,
and the Brown et al.~(2011) estimate depends upon complicated
planet-star interactions. If the planet has acted to spin-up the star
(recall the orbital plane and the stellar rotation axis are aligned),
the current relatively rapid stellar rotation rate would not be
reflective of true age. Stellar age has proven difficult to determine
in some other hot-Jupiter systems; for example, Schr{\"o}ter et
al.(2011) find that CoRoT-2a is \hbox{X-ray} bright and young but a
late-K companion is \hbox{X-ray} undetected, inconsistent with the
inferred system age, and a similar situation may apply for HD~189733
(Pillitteri et al.~2011). If WASP-18 were 2--3 times older than
current estimates, the intrinsic activity would still be somewhat low
but to a much less unusual degree. If the age is indeed correct, it is
natural to consider whether the extremely massive and close-in planet
could act to suppress, rather than enhance, stellar activity. The
tidal force exerted by WASP-18 upon its parent star is much greater
than is typical even for hot Jupiter systems, and in fact Arras et
al.~(2012) note that the small but apparently non-zero eccentricity
indicated by radial-velocity data is likely due to tidal fluid motion
on the star. Mid-type F stars have shallow outer convective zones;
perhaps the tidal pull is sufficient to repress dynamo activity in
WASP-18 or comparably extreme systems (of which none are, however,
currently known). This possibility could be assessed through
magnetohydrodynamic simulations.

\subsection{An additional challenge to observations of star-planet interaction}

Here we explore the possibility that a hotspot rotating upon the
stellar surface can potentially mimic a shorter-period signature of
star-planet interaction if only observed over a small fraction of the
stellar rotation. Figure~5 shows simulated examples for HD~179949 and
$\upsilon$~And, which have stellar/planetary periods of 7/3.09 and
14/4.618 days, respectively (Shkolnik et al.~2008). The half and full
sinusoid models for planet-induced activity chosen for these examples
are similar to those applied by Shkolnik et al.~(2005, 2008) to these
systems, but here for an edge-on inclination and with arbitrary
normalizations. (By construction the stellar hotspot amplitude is
taken as unity, and the relative amplitude of the interaction model is
then greater for HD~179949 than for $\upsilon$~And, also qualitatively
similar to the published models.) The probability of the stellar and
planetary phases co-aligning to mimic star-planet interaction for
these particular models is not very large (e.g., $\simeq$3\% of 10000
random phase offsets for each example yield total squared residuals
less than 0.1), but with additional freedom to adjust the relative
amplitudes, or to consider additional parameters (such as spot
latitude or system inclination, within a limited range) or to choose
from alternative model functional forms, or with a greater tolerance
for (apparent) outliers, this effect could potentially present a
significant source of contamination within a large sample of tested
stars. For the particular cases of HD~179949 and $\upsilon$~And it
must be emphasized that several observing runs were conducted, and in
some (but not all) of those runs variability of similar amplitude
similarly phased with the planetary orbit was observed (Shkolnik et
al.~2005, 2008; Poppenhaeger et al.~2011a), which clearly decreases
the probability of mistaking a stellar hotspot for one co-rotating
with the planet.

We note that, in general, searches for star-planet interaction based
on \ion{Ca}{2} core variability can guard against this type of false
positive at a given epoch through obtaining coverage over at least one
complete stellar rotation and over multiple planetary orbits (as was
done with our coverage of WASP-18). If searches over multiple epochs
find some instances in which core variability phases with the stellar
rotation and others in which it apparently matches the orbital period,
a comparison of the amplitudes can also check whether the observed
activity is likely to arise from distinct (intrinsic versus
planet-induced) sources. It would be odd if the activity in the
``off'' state (phased with stellar rotation rather than orbital
period) had a similar amplitude to that seen in the ``on'' state,
during which there is no obvious physical reason that the star should
otherwise go silent. It is not clear to us that the current data for
HD~179949 and $\upsilon$~And can definitively pass this test, but see,
e.g., Fares et al.~(2012). In any case, the conclusion is that
complete and extended coverage, preferably across multiple epochs and
multiple orbits per epoch, is essential for high-confidence detections
of star-planet interaction.

\begin{figure}
\includegraphics[scale=0.50]{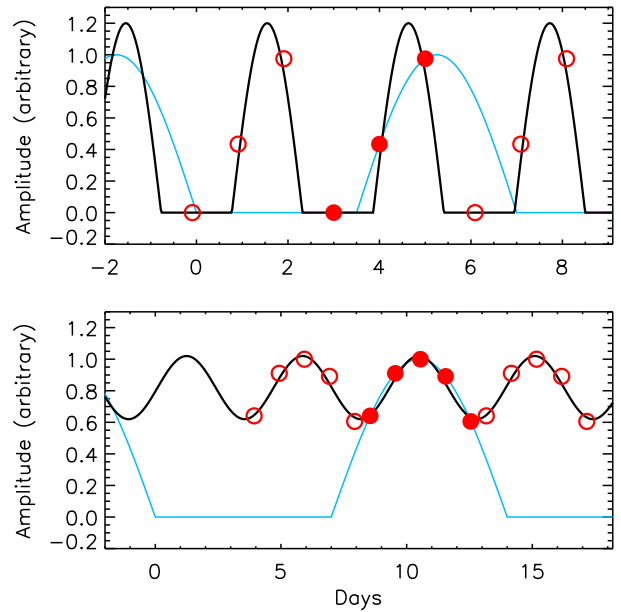} 
\figcaption{\small Examples of how a hotspot rotating with the star
(blue line) could potentially mimic various models of star-planet
interaction (black lines) over short observing runs (filled
circles). The stellar hotspot phased with the planetary orbit is shown
with open circles. See $\S$3.3 for discussion.}
\end{figure}

\section{Conclusions}

We have carried out a sensitive search for planet-induced stellar
activity within the extreme WASP-18 system, selected as an ideal
testbed for investigating potential magnetic (or tidal) interactions
between ``hot Jupiters'' and their host stars. High-resolution
spectroscopy of the \ion{Ca}{2} H and K lines was conducted with the
6.5m Magellan Clay Telescope and contemporaneous \hbox{X-ray}
monitoring was obtained with the {\it Swift\/} satellite. Our primary
results are the following:

1. The \ion{Ca}{2} H and K cores do not show significant variability
   over $\sim$8~d. Stacking residual spectra from 13 visits into phase
   bins of 0.70--0.93, 0.99--1.19, and 1.21--1.41 does not show any
   significant structural changes that could be attributed to
   planetary influence.

2. WASP-18 is not detected in a stacked 50~ks XRT exposure,
   constraining the \hbox{X-ray} luminosity to be unusually low for a
   young F6 star, with $\log{L_{\rm X}}<27.6$.

3. The lack of observed variability phased with the planetary orbit
   suggests any magnetic star-planet interaction in WASP-18 must
   be transient, if present at all.

4. The low level of chromospheric and coronal activity is consistent
   with an intrinsically weak magnetic field on WASP-18, perhaps
   indicating that the inferred young stellar age is not reliable, or
   alternatively potentially related to particularly strong planetary
   tidal effects.

5. It is demonstrated that a stellar hotspot can potentially mimic
   star-planet interaction, for observations truncated to a short
   segment of the rotation period.

6. Current ideas concerning star-planet interaction do not appear to
   be supported by the above results, therefore it may be optimistic
   at present to conceive of star-planet interaction as a robust
   estimator of exoplanet magnetic field strengths.

Further high-quality \ion{Ca}{2} H and K spectroscopy (as in Shkolnik
et al.~2005, 2008) of previously monitored and newly discovered
systems would refine understanding of the observational signatures of
star-planet interaction. In addition, we are currently carrying out a
{\it Chandra\/} survey of solar analogs to check whether stars with
close-in planets are systematically enhanced in \hbox{X-ray}
luminosity, and this experiment has been designed to sidestep many of
the selection biases that necessarily challenged previous large-sample
\hbox{X-ray} studies. That investigation will help further clarify
whether the difficulties in establishing observational evidence of
star-planet interaction in WASP-18 are anomalous or typical.

\acknowledgments 

We thank Mario Mateo for essential assistance with the Magellan
observing run, Neil Gehrels and the {\it Swift\/} Science Team for
approving and carrying out our Target of Opportunity observations of
WASP-18, and an anonymous referee for helpful comments. This paper
includes data gathered with the 6.5m Magellan Telescopes located at
Las Campanas Observatory, Chile. This research has made use of the
Exoplanet Orbit Database and the Exoplanet Data Explorer at
exoplanets.org. The Center for Exoplanets and Habitable Worlds is
supported by the Pennsylvania State University, the Eberly College of
Science, and the Pennsylvania Space Grant Consortium.


\begin{thebibliography}
\small 


\bibitem[Arge et al.(1995)]{1995ApJ...443..795A} Arge, C.~N., Mullan, 
   D.~J., \& Dolginov, A.~Z.\ 1995, \apj, 443, 795 

\bibitem[Arras et al.(2012)]{2012MNRAS.422.1761A} Arras, P., Burkart, J., 
  Quataert, E., \& Weinberg, N.~N.\ 2012, \mnras, 422, 1761 

\bibitem[Bodenheimer et al.(2001)]{2001ApJ...548..466B} Bodenheimer, P., 
   Lin, D.~N.~C., \& Mardling, R.~A.\ 2001, \apj, 548, 466 

\bibitem[Brown et al.(2011)]{2011MNRAS.415..605B} Brown, D.~J.~A.,
   Collier Cameron, A., Hall, C., Hebb, L., \& Smalley, B.\ 2011, \mnras,
   415, 605

\bibitem[Buccino \& Mauas(2008)]{2008A&A...483..903B} Buccino, A.~P., 
   \& Mauas, P.~J.~D.\ 2008, \aap, 483, 903 

\bibitem[Canto Martins et al.(2011)]{2011A&A...530A..73C} Canto
   Martins, B.~L., Das Chagas, M.~L., Alves, S., et al.\ 2011, \aap,
   530, A73

\bibitem[Cohen et al.(2009)]{2009ApJ...704L..85C} Cohen, O., Drake, J.~J., 
   Kashyap, V.~L., et al.\ 2009, \apjl, 704, L85 

\bibitem[Cohen et al.(2011)]{2011ApJ...733...67C} Cohen, O., Kashyap,
   V.~L., Drake, J.~J., et al.\ 2011, \apj, 733, 67

\bibitem[Cuntz et al.(2000)]{2000ApJ...533L.151C} Cuntz, M., Saar,
   S.~H., \& Musielak, Z.~E.\ 2000, \apjl, 533, L151

\bibitem[Fares et al.(2010)]{2010MNRAS.406..409F} Fares, R., Donati,
   J.-F., Moutou, C., et al.\ 2010, \mnras, 406, 409

\bibitem[Fares et al.(2012)]{2012MNRAS.423.1006F} Fares, R., Donati, J.-F., 
  Moutou, C., et al.\ 2012, \mnras, 423, 1006 

\bibitem[Freire Ferrero et al.(2004)]{2004A&A...413..657F} Freire Ferrero, R., 
   Frasca, A., Marilli, E., \& Catalano, S.\ 2004, \aap, 413, 657

\bibitem[Gu et al.(2005)]{2005AN....326..909G} Gu, P.-G., Shkolnik, E., Li, 
   S.-L., \& Liu, X.-W.\ 2005, Astronomische Nachrichten, 326, 909

\bibitem[Gurdemir et al.(2012)]{2012arXiv1202.3612G} Gurdemir, L., 
   Redfield, S., \& Cuntz, M.\ 2012, arXiv:1202.3612 

\bibitem[Hellier et al.(2009)]{2009Natur.460.1098H} Hellier, C.,
   Anderson, D.~R., Collier Cameron, A., et al.\ 2009, \nat, 460, 1098

\bibitem[Henry et al.(1996)]{1996AJ....111..439H} Henry, T.~J., Soderblom, 
   D.~R., Donahue, R.~A., \& Baliunas, S.~L.\ 1996, \aj, 111, 439

\bibitem[Kashyap et al.(2008)]{2008ApJ...687.1339K} Kashyap, V.~L.,
   Drake, J.~J., \& Saar, S.~H.\ 2008, \apj, 687, 1339

\bibitem[Kraft et al.(1991)]{1991ApJ...374..344K} Kraft, R.~P., Burrows, 
   D.~N., \& Nousek, J.~A.\ 1991, \apj, 374, 344 

\bibitem[Krej{\v c}ov{\'a} \& Budaj(2012)]{2012A&A...540A..82K} 
   Krej{\v c}ov{\'a}, T., \& Budaj, J.\ 2012, \aap, 540, A82 

\bibitem[Lanza(2008)]{2008A&A...487.1163L} Lanza, A.~F.\ 2008, \aap,
   487, 1163

\bibitem[Lanza(2009)]{2009A&A...505..339L} Lanza, A.~F.\ 2009, \aap, 
   505, 339 

\bibitem[Lanza et al.(2011)]{2011A&A...525A..14L} Lanza, A.~F.,
   Bonomo, A.~S., Pagano, I., et al.\ 2011, \aap, 525, A14

\bibitem[Lenz et al.(2010)]{2010arXiv1012.1720L} Lenz, L.~F., Reiners,
   A., K{\"u}rster, M.\ 2011, in ASP Conf. Ser. 448, 16th Cambridge
   Workshop on Cool Stars, Stellar Systems, and the Sun,
   ed.~C.~Johns-Krull (San Francisco, CA: ASP), 1173

\bibitem[Mamajek \& Hillenbrand(2008)]{2008ApJ...687.1264M} 
   Mamajek, E.~E., \& Hillenbrand, L.~A.\ 2008, \apj, 687, 1264 

\bibitem[Moretti et al.(2005)]{2005SPIE.5898..360M} Moretti, A., Campana, 
   S., Mineo, T., et al.\ 2005, \procspie, 5898, 360

\bibitem[Nymeyer et al.(2011)]{2011ApJ...742...35N} Nymeyer, S.,
   Harrington, J., Hardy, R.~A., et al.\ 2011, \apj, 742, 35

\bibitem[Pevtsov et al.(2003)]{2003ApJ...598.1387P} Pevtsov, A.~A., Fisher, 
   G.~H., Acton, L.~W., et al.\ 2003, \apj, 598, 1387

\bibitem[Pillitteri et al.(2010)]{2010ApJ...722.1216P} Pillitteri, I.,
   Wolk, S.~J., Cohen, O., et al.\ 2010, \apj, 722, 1216

\bibitem[Pillitteri et al.(2011)]{2011ApJ...741L..18P} Pillitteri, I., 
   G{\"u}nther, H.~M., Wolk, S.~J., Kashyap, V.~L.,
   \& Cohen, O.\ 2011, \apjl, 741, L18 

\bibitem[Poppenhaeger et al.(2009)]{2009A&A...508.1417P}
   Poppenh{\"a}ger, K., Robrade, J., Schmitt, J.~H.~M.~M., \& Hall,
   J.~C.\ 2009, \aap, 508, 1417

\bibitem[Poppenhaeger et al.(2010)]{2010A&A...515A..98P}
   Poppenh{\"a}ger, K., Robrade, J., \& Schmitt, J.~H.~M.~M.\ 2010, \aap,
   515, A98

\bibitem[Poppenhaeger et al.(2011)]{2011A&A...528A..58P}
   Poppenh{\"a}ger, K., Lenz, L.~F., Reiners, A., Schmitt, J.~H.~M.~M.,
   \& Shkolnik, E.\ 2011a, \aap, 528, A58

\bibitem[Poppenhaeger \& Schmitt(2011)]{2011ApJ...735...59P}
   Poppenh{\"a}ger, K., \& Schmitt, J.~H.~M.~M.\ 2011b, \apj, 735, 59

\bibitem[Poppenhaeger et al.(2012)]{2012AN....333...26P} Poppenhaeger, K., 
   G{\"u}nther, H.~M., \& Schmitt, J.~H.~M.~M.\ 2012, 
   Astronomische Nachrichten, 333, 26 

\bibitem[Preibisch \& Feigelson(2005)]{2005ApJS..160..390P} 
   Preibisch, T., \& Feigelson, E.~D.\ 2005, \apjs, 160, 390

\bibitem[Rubenstein \& Schaefer(2000)]{2000ApJ...529.1031R}
   Rubenstein, E.~P., \& Schaefer, B.~E.\ 2000, \apj, 529, 1031

\bibitem[Saar et al.(2008)]{2008IAUS..249...79S} Saar, S.~H., Cuntz,
   M., Kashyap, V.~L., \& Hall, J.~C.\ 2008, IAU Symposium, 249, 79

\bibitem[S{\'a}nchez-Lavega(2004)]{2004ApJ...609L..87S} S{\'a}nchez-Lavega, 
   A.\ 2004, \apjl, 609, L87

\bibitem[Scharf(2010)]{2010ApJ...722.1547S} Scharf, C.~A.\ 2010, \apj,
   722, 1547

\bibitem[Schmitt(1997)]{1997A&A...318..215S} Schmitt, J.~H.~M.~M.\ 
   1997, \aap, 318, 215 

\bibitem[Schmitt \& Liefke(2004)]{2004A&A...417..651S} Schmitt, 
   J.~H.~M.~M., \& Liefke, C.\ 2004, \aap, 417, 651

\bibitem[Schr{\"o}ter et al.(2011)]{2011A&A...532A...3S} 
   Schr{\"o}ter, S., Czesla, S., Wolter, U., et al.\ 2011, \aap, 532, A3 

\bibitem[Shkolnik et al.(2005)]{2005ApJ...622.1075S} Shkolnik, E.,
   Walker, G.~A.~H., Bohlender, D.~A., Gu, P.-G., K{\"u}rster, M.\ 2005,
   \apj, 622, 1075

\bibitem[Shkolnik et al.(2008)]{2008ApJ...676..628S} Shkolnik, E.,
   Bohlender, D.~A., Walker, G.~A.~H., \& Collier Cameron, A.\ 2008,
   \apj, 676, 628

\bibitem[Southworth et al.(2009)]{2009ApJ...707..167S} Southworth, J.,
   Hinse, T.~C., Dominik, M., et al.\ 2009, \apj, 707, 167

\bibitem[Stelzer \& Neuh{\"a}user(2001)]{2001A&A...377..538S} 
   Stelzer, B., \& Neuh{\"a}user, R.\ 2001, \aap, 377, 538

\bibitem[Stevens(2005)]{2005MNRAS.356.1053S} Stevens, I.~R.\ 2005, \mnras, 
   356, 1053

\bibitem[Triaud et al.(2010)]{2010A&A...524A..25T} Triaud, A.~H.~M.~J., 
   Collier Cameron, A., Queloz, D., et al.\ 2010, \aap, 524, A25

\bibitem[Walker et al.(2008)]{2008A&A...482..691W} Walker, G.~A.~H.,
   Croll, B., Matthews, J.~M., et al.\ 2008, \aap, 482, 691

\bibitem[Winn et al.(2010)]{2010ApJ...718L.145W} Winn, J.~N., Fabrycky, D., 
   Albrecht, S., \& Johnson, J.~A.\ 2010, \apjl, 718, L145 

\bibitem[Wright et al.(2004)]{2004ApJS..152..261W} Wright, J.~T.,
   Marcy, G.~W., Butler, R.~P., \& Vogt, S.~S.\ 2004, \apjs, 152, 261

\bibitem[Wright et al.(2011)]{2011PASP..123..412W} Wright, J.~T.,
   Fakhouri, O., Marcy, G.~W., et al.\ 2011, \pasp, 123, 412


\normalsize
\end{thebibliography}
\end{document}